\documentclass[aps,onecolumn,preprintnumbers,nofootinbib,superscriptaddress]{revtex4}
\usepackage{amsmath}
\usepackage{graphicx}
\usepackage{amsfonts}
\usepackage{array}
\usepackage{amsthm}
\usepackage{bm}
\usepackage{palatino}
\usepackage{mathpazo}
\usepackage{supertabular}
\usepackage{subfig}
\usepackage[breaklinks]{hyperref}
\usepackage{color}
\usepackage[font=small,labelfont=bf,
justification=justified,
format=plain]{caption}
\usepackage{graphicx}
\usepackage{caption}
\usepackage[font=small,labelfont=bf,justification=justified]{caption}
\usepackage[labelfont=bf,labelsep=quad,justification=justified]{caption}
\usepackage{epstopdf}
\usepackage[compat=1.1.0]{tikz-feynman}
\usetikzlibrary{calc,decorations.markings,decorations.pathmorphing,
	trees,positioning,arrows}
\providecommand{\keywords}[1]
{
	\small	
	\textbf{\textit{Keywords:}} #1
}
\begin{document}

\title{Geodesic motion of a test particle around a noncommutative Schwarzchild Anti-de Sitter black hole}

\author{Mohamed Aimen Larbi}
\email{mohamedaimen.larbi@univ-batna.dz}
\affiliation{Laboratory of Physics of Radiation and its Interactions with Matter (LRPRIM), University of Batna-1, Batna 05000, Algeria }
\affiliation{Department of Physics, Faculty of Matter Sciences, University of Batna-1, Batna 05000, Algeria}
\author{Slimane Zaim}
\email{zaim69slimane@yahoo.com}
\affiliation{Department of Physics, Faculty of Matter Sciences, University of Batna-1, Batna 05000, Algeria}
\author{Abdellah Touati}
\email{touati.abph@gmail.com}
\affiliation{Department of Physics, Faculty of Sciences and Applied Sciences, University of Bouira, Algeria}

\begin{abstract}
In this work, we derive non-commutative corrections to the Schwarzschild-Anti-de Sitter solution up to the first and second orders of the non-commutative parameter $\Theta$. Additionally, we obtain the corresponding deformed effective potentials and the non-commutative geodesic equations for massive particles. Through the analysis of time-like non-commutative geodesics for various values of $\Theta$, we demonstrate that the circular geodesic orbits of the non-commutative Schwarzschild-Anti-de Sitter black hole exhibit greater stability compared to those of the commutative one. Furthermore, we derive corrections to the perihelion deviation angle per revolution as a function of $\Theta$. By applying this result to the perihelion precession of Mercury and utilizing experimental data, we establish a new upper bound on the non-commutative parameter, estimated to be on the order of $10^{-66}\,\mathrm{m}^2$.
\end{abstract}


\date{\today}
\maketitle

\section{Introduction}

General Relativity (GR) is widely regarded as the most successful theory of gravity, providing highly accurate descriptions of various phenomena within the solar system, including the perihelion shift of Mercury, the deflection of light, and gravitational time delays, among others \cite{weinberg}. These results stem from analyzing the geodesic motion of celestial bodies around compact objects, a critical aspect of astrophysics that aids in understanding the geometry and dynamics of gravitational systems. In this context, extensive research has been conducted on the geodesic motion of different types of particles around various compact objects, as explored in Refs. \cite{Chandr1,hackmann1,kagraman,hackmann3,levin,grunau,belbrun,barack,pugliese1,pugliese2,virb1,virb2,pradhan,pugl1,pugl2}.

Despite its remarkable success, GR encounters significant challenges in explaining certain astrophysical phenomena, such as galactic rotation curves, gravitational lensing, the structure of the cosmic microwave background, and the accelerated expansion of the universe. These issues have led to the introduction of two enigmatic concepts: Dark Matter and Dark Energy. While these concepts provide partial explanations, they remain fundamentally mysterious, prompting the development of numerous theories that extend beyond GR.

One notable extension is the inclusion of the cosmological constant in GR, which provides a successful model for describing Dark Energy (non-zero vacuum energy) and explains the observed accelerated expansion of the universe \cite{cosmological1}. Recent studies have focused on both repulsive ($\Lambda > 0$) \cite{cosmological2} and attractive ($\Lambda < 0$) \cite{cosmological3} cosmological constants, described by vacuum solutions of Einstein’s equations corresponding to de Sitter (dS) and anti-de Sitter (AdS) Schwarzschild black holes (BHs), respectively. Significant attention has also been given to geodesic motion around BHs in the presence of a cosmological constant \cite{abd3,abd4,abd5,abd6,abd7,abd11,abd12,1110.3,1110.5,cosmological4,cosmological5,islam1,jaklit,stuchli,gibbons1,kran1,kran2,hackmann2,olivares,villanueva}.

Our objective is to investigate the geodesic motion of a massive test particle around an AdS Schwarzschild BH in non-commutative (NC) spacetime. Specifically, we aim to derive the geodesic equation from a metric modified by NC geometry using the NC product between coordinates. We also perform a detailed analysis of the NC-corrected effective potentials and examine the influence of NC parameters on the stability of circular orbits. Moreover, we compute NC corrections to the perihelion deviation angle per revolution, applying our results to Mercury’s perihelion shift. Using experimental data, we estimate a value for the NC parameter $\Theta$, finding it close to the length of Planck's constant. NC geometry, which naturally emerges from string theory \cite{1110.9,berger}, offers a promising framework for quantizing gravity at quantum scales \cite{noncommutative1}. This framework posits that gravity quantization can be achieved by quantizing spacetime itself, predicting a fundamental minimal length at the Planck scale. Our study explores the implications of this fundamental length on macroscopic astrophysical phenomena \cite{abdellah6,abdellah4,abdellah5}.

NC geometry traces its origins to the works of Weil and Moyal, who studied phase-space quantization \cite{weyl}. Snyder \cite{snyder1,snyder2} later developed a consistent theory of NC coordinate space based on algebraic representations. Over the past few decades, NC physics has garnered increasing interest, driven by string theory, quantum gravity \cite{KALAU,kastler,chamseddine}, the Standard Model \cite{CONNES,VARILLY,martin,szabo,seiberg}, and the quantum Hall effect \cite{bellissard}. In NC spacetime, the coordinates $\hat{x}^\mu$ satisfy the commutation relation:
\begin{equation}
	\left[ \hat{x}^\mu, \hat{x}^\nu \right] = i \Theta^{\mu \nu},
	\label{eq1}
\end{equation}
where $\Theta^{\mu \nu}$ is a constant anti-symmetric tensor with dimensions of $(\text{length})^2$. A relation between the ordinary coordinate system and the NC coordinate system is given by the Bopp shift:
\begin{equation}
	\hat{x}^\mu = x^\mu - \frac{1}{2} \overline{\Theta}^{\mu \nu} p_\nu,
	\label{eq2}
\end{equation}
where $\overline{\Theta}^{\mu \nu} = \Theta^{\mu \nu} / \hbar$. The ordinary product is replaced by the Moyal product \cite{snyder1}:
\begin{equation}
	(f \ast g)(x) = f(x) e^{\frac{i}{2} \Theta^{\mu \nu} \overleftarrow{\partial_\mu} \overrightarrow{\partial_\nu}} g(x).
	\label{eq4}
\end{equation}

Recent studies have explored the effects of NC geometry on geodesic equations, both generally \cite{nozari1,nozari2,kuniyal,mirza,rome1,iftikhar,ulh1,nicolini1,bhar,rahaman,abd,abdellah3,abdellah2} and specifically in the presence of a cosmological constant \cite{abd41}. Our study contributes to this body of work by deriving the geodesic equation from the NC-deformed AdS Schwarzschild metric. We also compute NC corrections to the effective potential and the perihelion shift, estimating $\sqrt{\Theta}$ to be approximately $10^{-33} \, \mathrm{m}$. Finally, we discuss the stability of circular orbits in NC AdS Schwarzschild spacetime.

This article is organized as follows: In Sect. \ref{sec:NCADSSBH}, we present the NC corrections to the AdS Schwarzschild metric using Bopp’s shift. In Sect. \ref{sec:NCEPSC}, we derive the NC-corrected effective potentials and discuss the stability of circular orbits. In Sect. \ref{sec:NCOEAP}, we compute NC corrections to the geodesic equation and the perihelion shift, providing experimental constraints on $\Theta$. Finally, Sect. \ref{sec:conl} summarizes our conclusions.


\section{Non-Commutative Schwarzschild-AdS Black Hole}\label{sec:NCADSSBH}

The spherically symmetric Schwarzschild--de Sitter BH is the simplest solution to the Einstein field equations with a cosmological constant: $R_{\mu\nu} - \frac{1}{2}R g_{\mu\nu} + \Lambda g_{\mu\nu} = 0,$
describing an isolated mass in a spherically symmetric spacetime. This model has been highly successful in explaining the observed accelerated expansion of the universe \cite{weinberg}. The metric of the AdS Schwarzschild spacetime is given by (assuming $c = G = 1$):
\begin{equation} 
	ds^2 = -\left(1 - \frac{2m}{r} - \frac{\Lambda}{3}r^2\right)dt^2 + \left(1 - \frac{2m}{r} - \frac{\Lambda}{3}r^2\right)^{-1}dr^2 + r^2\left(d\theta^2 + \sin^2\theta \, d\varphi^2\right),
\end{equation}
where $m$ represents the Schwarzschild mass, and $\Lambda$ is the negative cosmological constant.

To obtain the deformed components of the AdS Schwarzschild metric in a NC spacetime, we apply Bopp's shift transformation to the coordinates, as outlined in Eq.~\eqref{eq2}. We consider only spatial non-commutativity ($\overline{\Theta}^{0i} = 0$) to avoid known issues with unitarity. To simplify calculations, we choose the antisymmetric tensor for the non-commutativity parameter as:
\begin{equation}
	\overline{\Theta }^{\mu \nu }= 
	\begin{pmatrix}
		0 & 0 & 0 & 0 \\
		0 & 0 & 0 & \overline{\Theta} \\
		0 & 0 & 0 & 0 \\
		0 & -\overline{\Theta } & 0 & 0 \\
	\end{pmatrix}, 
	\hspace{0.5cm} \mu,\nu = 0,1,2,3.
	\label{eq6}
\end{equation}

Using this in Eq.~\eqref{eq2}, the NC coordinates transform as:
\begin{align}
	\hat{r} &= r - \frac{1}{2}\overline{\Theta} p_\varphi, \label{eq7} \\
	\hat{\varphi} &= \varphi + \frac{1}{2}\overline{\Theta} p_r, \label{eq8}
\end{align}
where $p_\varphi$ and $p_r$ are the momentum components along the $\varphi$ and $r$ directions, respectively. Substituting Eq.~\eqref{eq7} into the metric, the NC Schwarzschild-AdS spacetime metric $\hat{g}_{\mu \nu}$ up to the first order in $\overline{\Theta}$ takes the form:
\begin{equation}
	ds^2 = \hat{g}_{tt}(r,\overline{\Theta})dt^2 + \hat{g}_{rr}(r,\overline{\Theta})dr^2 + \hat{g}_{\theta\theta}(r,\overline{\Theta})d\theta^2 + \hat{g}_{\varphi\varphi}(r,\overline{\Theta})d\varphi^2,
	\label{eq9}
\end{equation}
where the components of the deformed metric up to first order are:
\begin{align} 
	-\hat{g}_{tt} &= \left(1 - \frac{2m}{r} - \frac{\Lambda}{3}r^2\right) - \overline{\Theta} p_\varphi \left\{\frac{m}{r^2} - \frac{\Lambda}{3}r\right\} + \mathcal{O}(\overline{\Theta}^2),
	\label{eq10} \\
	\hat{g}_{rr} &= \left(1 - \frac{2m}{r} - \frac{\Lambda}{3}r^2\right)^{-1} + \overline{\Theta} p_\varphi \left\{\left(\frac{m}{r^2} - \frac{\Lambda}{3}r\right)\left(1 - \frac{2m}{r} - \frac{\Lambda}{3}r^2\right)^{-2}\right\} + \mathcal{O}(\overline{\Theta}^2),
	\label{eq11} \\    
	\hat{g}_{\theta\theta} &= r^2 - r \overline{\Theta} p_\varphi + \mathcal{O}(\overline{\Theta}^2),
	\label{eq12} \\    
	\hat{g}_{\varphi\varphi} &= r^2\sin^2\theta - r \overline{\Theta} p_\varphi \sin^2\theta + \mathcal{O}(\overline{\Theta}^2).
	\label{eq13}
\end{align}

Notably, after rearrangement, it can be shown that $\hat{g}_{tt} = 1/\hat{g}_{rr}$. In the limit $\overline{\Theta}\rightarrow 0$, the metric reduces to the standard Schwarzschild-AdS BH solution \cite{Ads.Sch}. Furthermore, for $\Lambda = 0$, the metric reduced to the NC Schwarzschild BH \cite{nasseri}.

The NC event horizon is determined by solving $\hat{g}_{tt} = 0$, leading to the event horizon radius at first order in $\overline{\Theta}$:
\begin{equation}
	r_h^{NC} = 2m + \frac{8\Lambda m^3}{3} + \frac{1}{2}\overline{\Theta} p_\varphi.
	\label{eq14}
\end{equation}
When $\overline{\Theta} = 0$, the commutative event horizon radius is recovered.

\begin{figure}[ht]
		\includegraphics[width=0.5\linewidth]{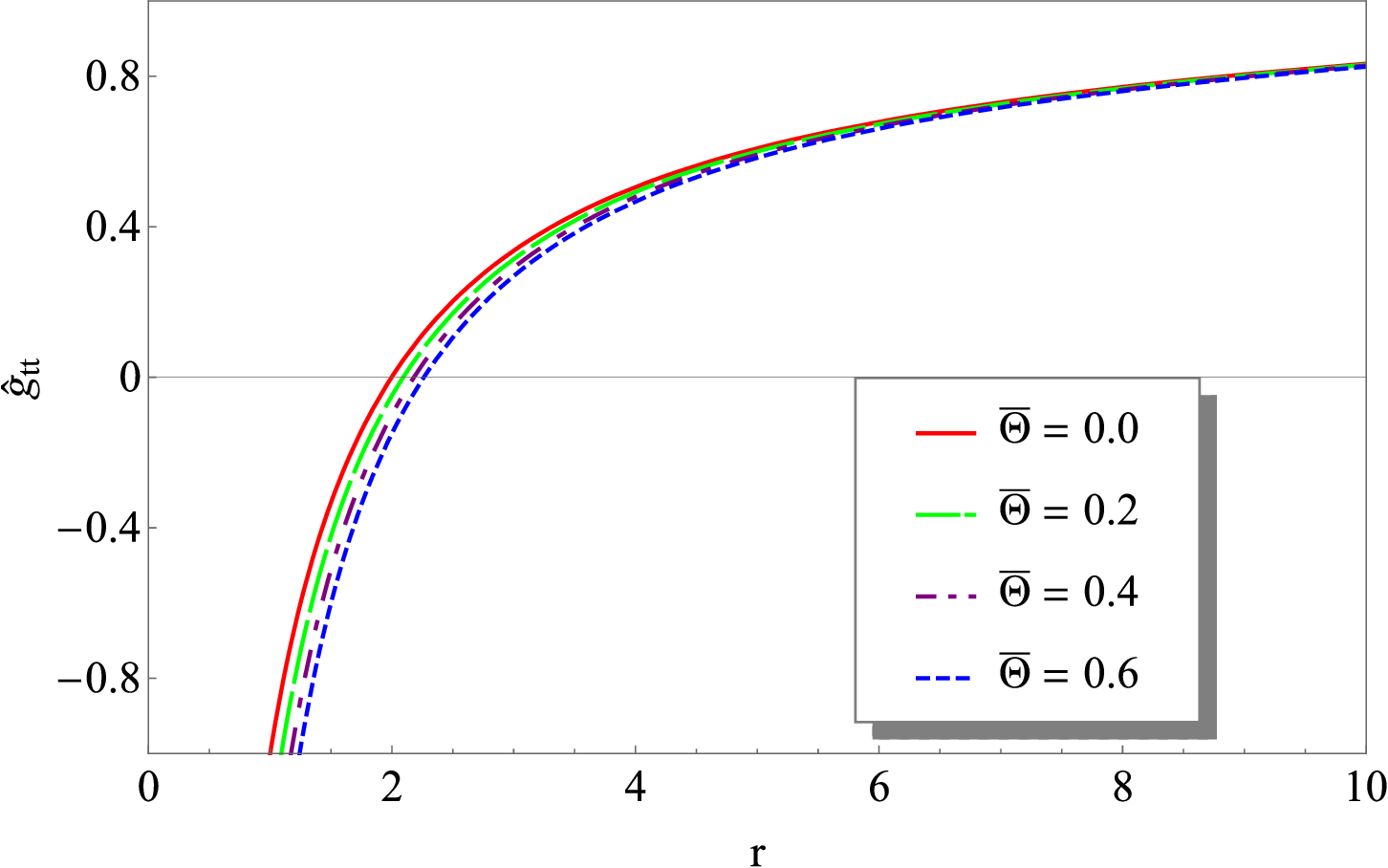}\hfill
		\includegraphics[width=0.5\linewidth]{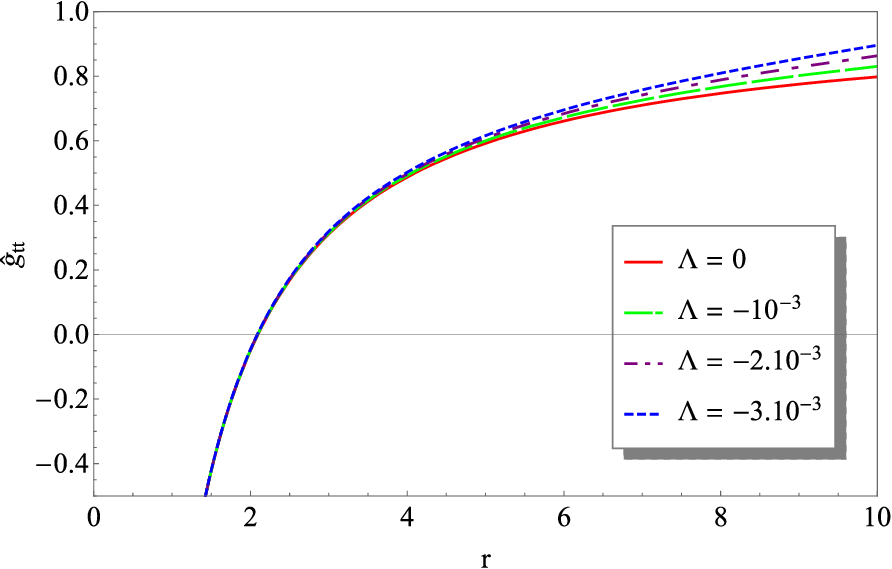}
	\caption{The behavior of $\hat{g}_{tt}$ in NC spacetime. \textbf{Left:} Varying $\overline{\Theta}$ with $m = 1$, $\Lambda = -10^{-3}$, and $p_\varphi = 1$. \textbf{Right:} Varying $\Lambda$ with $m = 1$, $p_\varphi = 1$, and $\overline{\Theta} = 0.2$.}
	\label{fig1}
\end{figure}

Figure~\ref{fig1} demonstrates the influence of non-commutativity and the cosmological constant on $\hat{g}_{tt}$. The left panel illustrates how increasing $\overline{\Theta}$ increases the event horizon radius ($r_h^{NC} > r_h$), an effect that decreases with distance. The right panel shows that the cosmological constant significantly impacts $\hat{g}_{tt}$ at large distances, while its influence near the event horizon is negligible.

\begin{figure}[ht]
	\centering
	\includegraphics[width=0.6\textwidth]{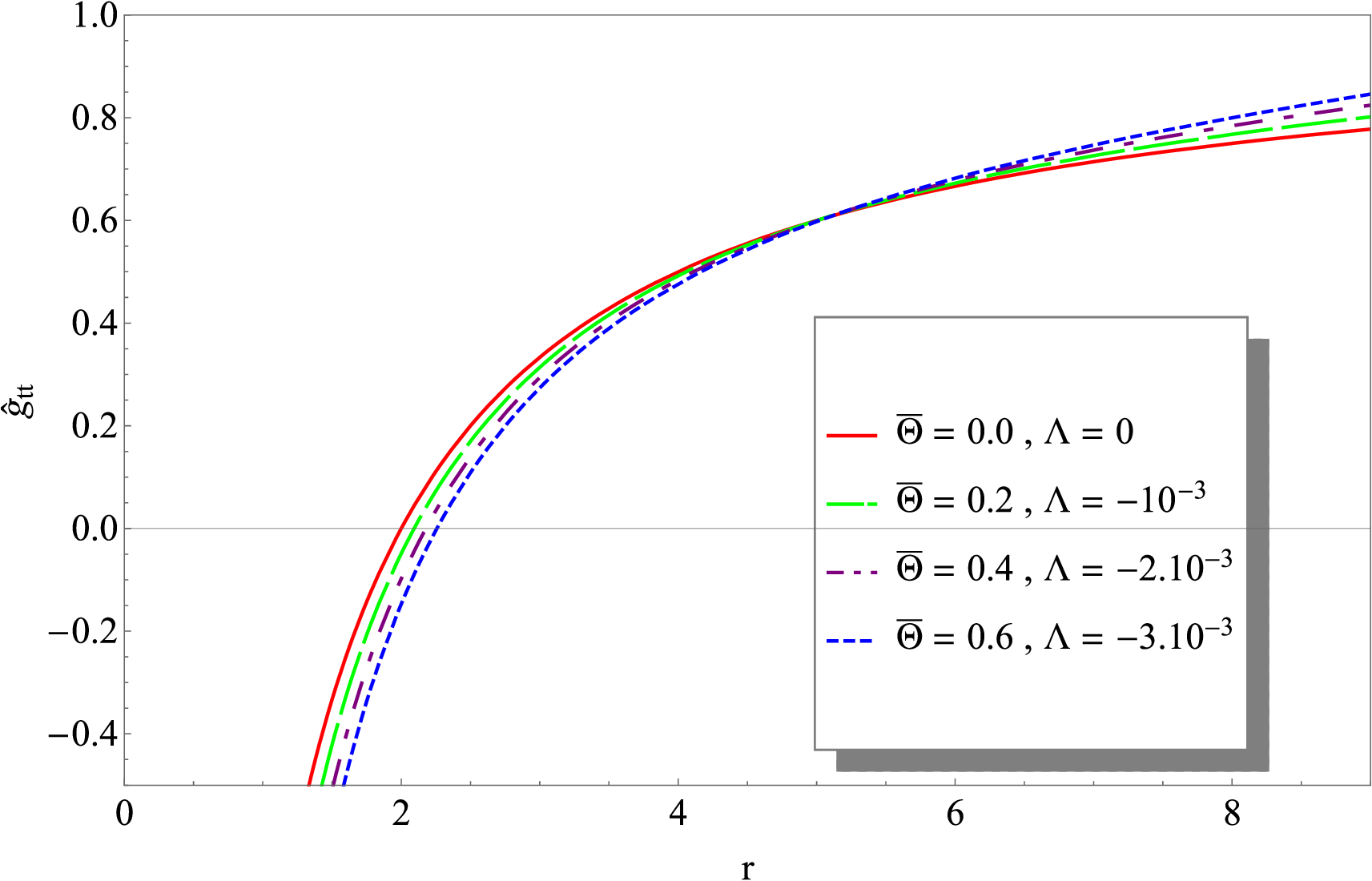}
	\caption{The behavior of $\hat{g}_{tt}$ in NC spacetime for varying $\overline{\Theta}$ and $\Lambda$, with $m = 1$ and $p_\varphi = 1$.}
	\label{fig2}
\end{figure}

Figure \ref{fig2} combines the results presented in Figure \ref{fig1}, illustrating the complementary effects of NC geometry and the cosmological constant. The influence of NC geometry decrease with distance from the event horizon, while the effect of the cosmological constant becomes increasingly relevant at larger distances. The graph also clearly identifies the transition point where the impact of NC geometry vanishes, and the cosmological constant begins to dominate. This distinction arises due to their different physical origins. Non-commutativity arises from the high-density matter near the BH singularity, where quantum gravitational effects necessitate a fundamental limit on position measurements. This limit effectively replaces the classical notion of spacetime points with Planck-scale cells, making NC geometry a useful framework for describing physics in regions of extreme density, such as near the singularity or during the early stages of the universe (e.g., the Big Bang).
In contrast, the cosmological constant represents an external energy source attributed to the vacuum energy of spacetime. It is often associated with dark energy and is instrumental in explaining the accelerated expansion of the universe. While NC geometry primarily influences small-scale, high-density regions near the black hole, the cosmological constant governs large-scale dynamics, becoming increasingly relevant at cosmological distances.


\section{Geodesics in the NC Schwarzschild-AdS metric}\label{sec:NCEPSC}

In general relativity, geodesics describe the trajectories of test particles in spacetime and encompass various structures, including those in regular black hole spacetimes. The behavior of these geodesics is governed by two constants of motion: the energy and angular momentum of the particles. By parametrizing these constants, one can classify the different types of trajectories around a black hole, such as those where particles fall into the black hole, scatter away, or form stable or unstable orbits. The specific trajectory depends on the particle's initial conditions, particularly its angular momentum and energy. 

In this work, we extend the study of geodesics to a NC Schwarzschild-AdS spacetime. In this scenario, the NC geometry introduces a new constant $\overline{\Theta}$ in the geodesic equation of test particle around this BH, in addition to the two constants of motion complements the energy and angular momentum. To analyze the geodesic structure in the NC Schwarzschild-AdS spacetime described by Eq.~\eqref{eq9}, we restrict our investigation to the equatorial plane ($\theta = \pi / 2$). In this plane, the equations of motion are derived by solving the Euler-Lagrange equations for the variational problem associated with the spacetime metric. The corresponding Lagrangian is expressed as:
\begin{equation}
	2\mathcal{L} = \hat{g}_{tt}(r, \overline{\Theta}) \dot{t}^2 + \hat{g}_{rr}(r, \overline{\Theta}) \dot{r}^2 + \hat{g}_{\varphi\varphi}(r, \overline{\Theta}) \dot{\varphi}^2
	\label{eq15}
\end{equation}

Since the Lagrangian is independent of $t$ and $ \varphi $, there are two constants of motion:
\begin{subequations}
	\begin{align}
		E &= \frac{\partial \mathcal{L}}{\partial \dot{t}} = \hat{g}_{tt}(r, \overline{\Theta}) \dot{t} \quad \Rightarrow \quad \dot{t} = \frac{E}{\hat{g}_{tt}(r, \overline{\Theta})}
		\label{eq16a} \\
		L &= \frac{\partial \mathcal{L}}{\partial \dot{\varphi}} = \hat{g}_{\varphi\varphi}(r, \overline{\Theta}) \dot{\varphi} \quad \Rightarrow \quad \dot{\varphi} = \frac{L}{\hat{g}_{\varphi\varphi}(r, \overline{\Theta})}
		\label{eq16b}
	\end{align}
\end{subequations}
where $ E $ and $ L $ are the energy and angular momentum of the test particle, respectively.

Using the condition $ \hat{g}_{\mu \nu} \frac{dx^{\mu}}{d\tau} \frac{dx^{\nu}}{d\tau} = h $, where $h = 1$ or $ h = 0 $ for a massive or massless particle, the geodesic equation reduces to a single NC differential equation:
\begin{equation}
	\dot{r}^2 = E^2 - V_{\text{eff}}^2(r, \overline{\Theta})
	\label{eq17}
\end{equation}
where the NC effective potential is given by:
\begin{equation}
	\left( V_{\text{eff}}^{\text{NC}}\right) ^2(r, \overline{\Theta}) = \left(1 - \frac{2M}{r} - \frac{\Lambda}{3} r^2\right)\left(\frac{L^2}{r^2} + h\right) - \overline{\Theta} p_\varphi \left\{\frac{3L^2 m}{r^4} - \frac{L^2}{r^3} + \frac{h m}{r^2} - h \frac{\Lambda}{3} r\right\} + \mathcal{O}(\overline{\Theta}^2)
	\label{eq18}
\end{equation}
In the limit $ \overline{\Theta} \to 0 $, we recover the commutative result \cite{cosmological5}.

\begin{figure}[ht]
	\centering
	\includegraphics[width=0.48\textwidth]{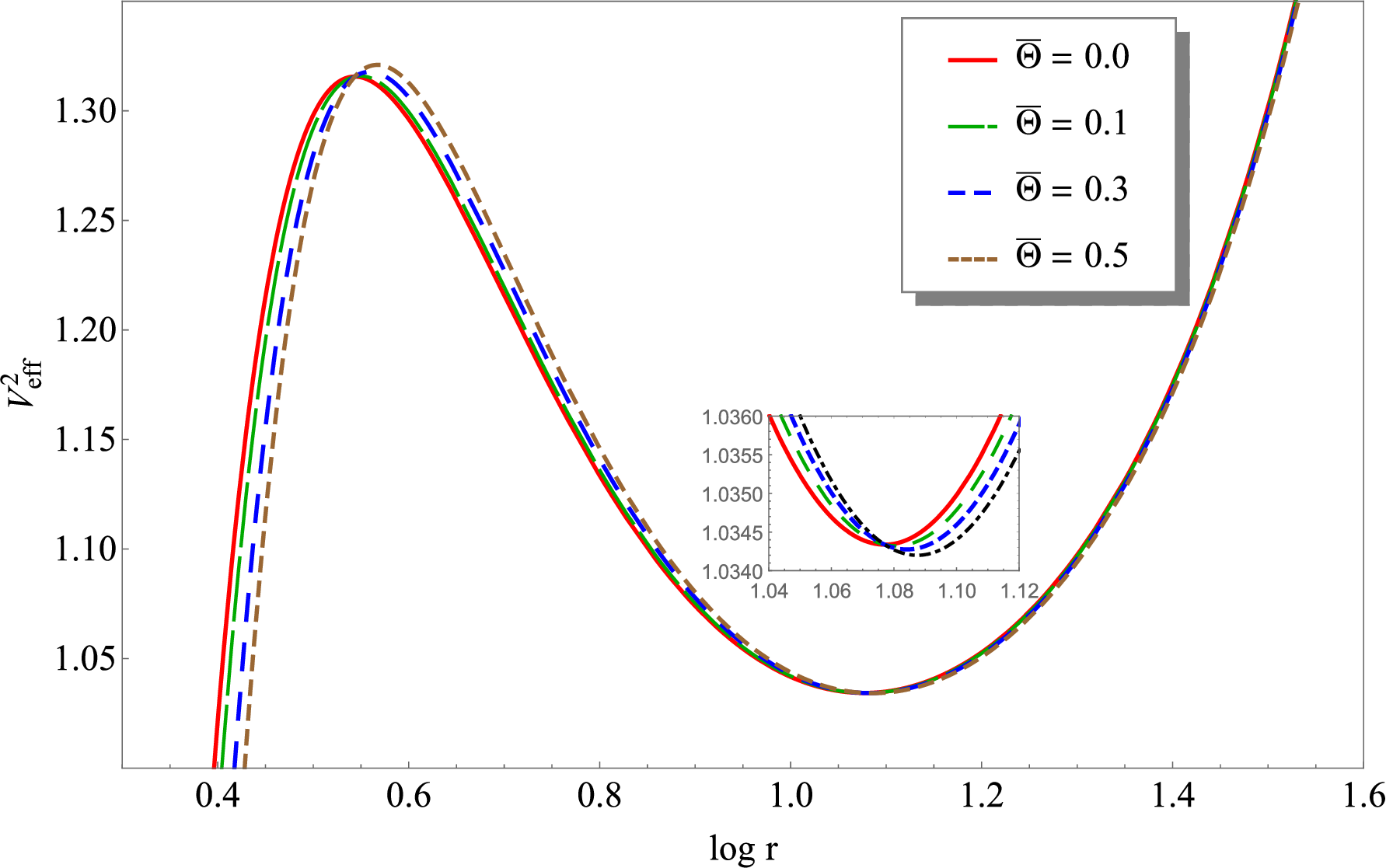}
	\includegraphics[width=0.48\textwidth]{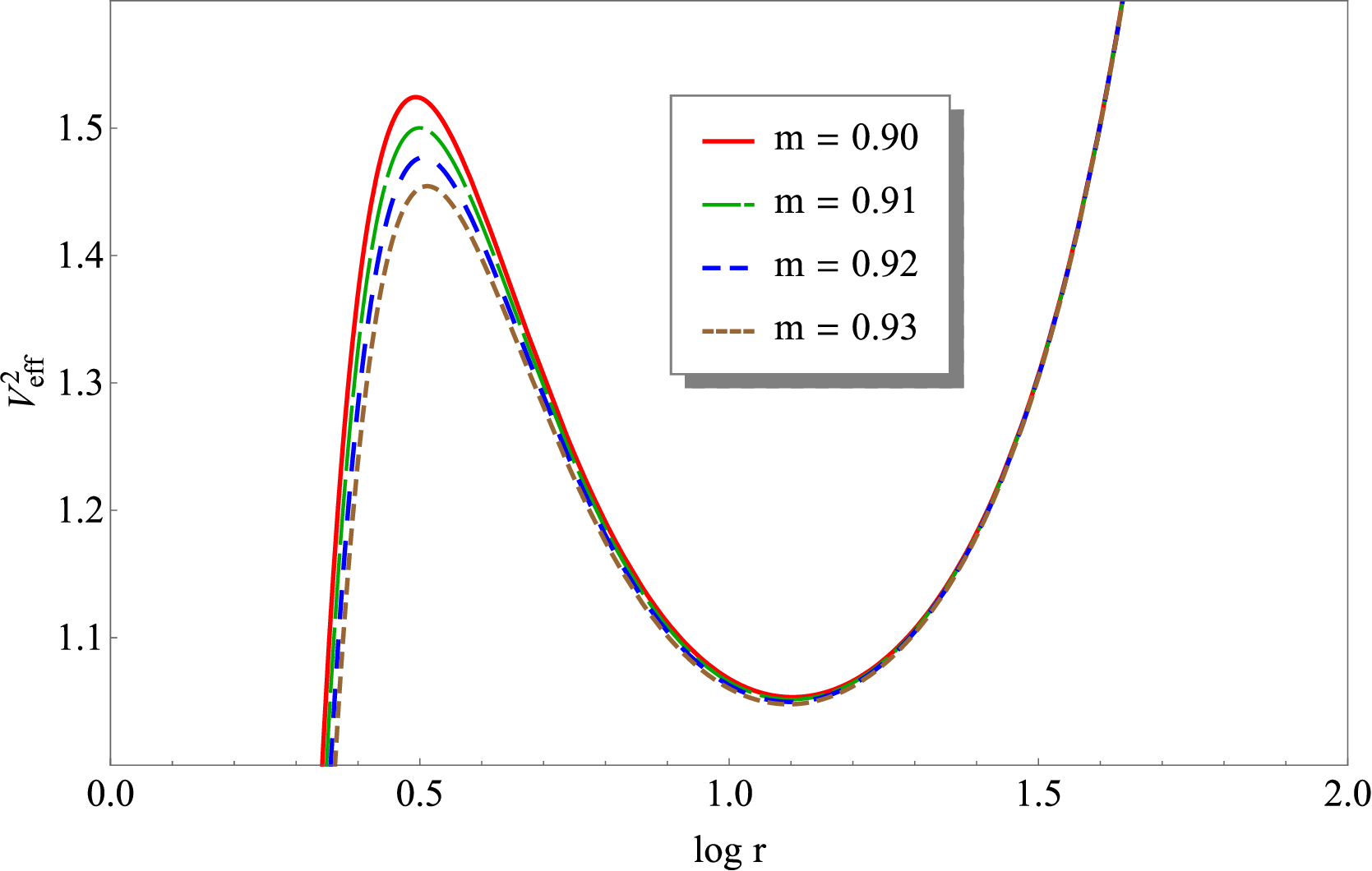}
	\caption{The behavior of the effective potential for a massive test particle in NC AdS Schwarzschild spacetime. \textbf{Left panel:} Different values of $\overline{\Theta}$, with fixed $ m = 1$, $ \Lambda = -10^{-3} $, $ p_\varphi = 1 $, and $ L = 5 $. \textbf{Right panel:} Different values of $ m $ and $ p_\varphi $, with fixed $ \Lambda = -10^{-3} $, $ \overline{\Theta} = 0.2 $, and $L = 5$.}
	\label{fig3}
\end{figure}

Figure \ref{fig3} shows the influence of the parameters $\overline{\Theta}$ and $m$ on the NC effective potential for a massive test particle. In the left panel, as $ \overline{\Theta} $ increases, the peak of the effective potential rises and shifts slightly outward from the event horizon. The influence of $\overline{\Theta}$ decreases at larger distances. In the right panel, as $m$ increases, the peak of the NC effective potential decreases and moves farther from the event horizon, and the depth of the potential well decreases similarly. We observe that in the NC spacetime, the effective potential $V_{\text{eff}}^{\text{NC}}$ becomes more stable.

\begin{figure}[ht]
	\centering
	\includegraphics[width=0.48\textwidth]{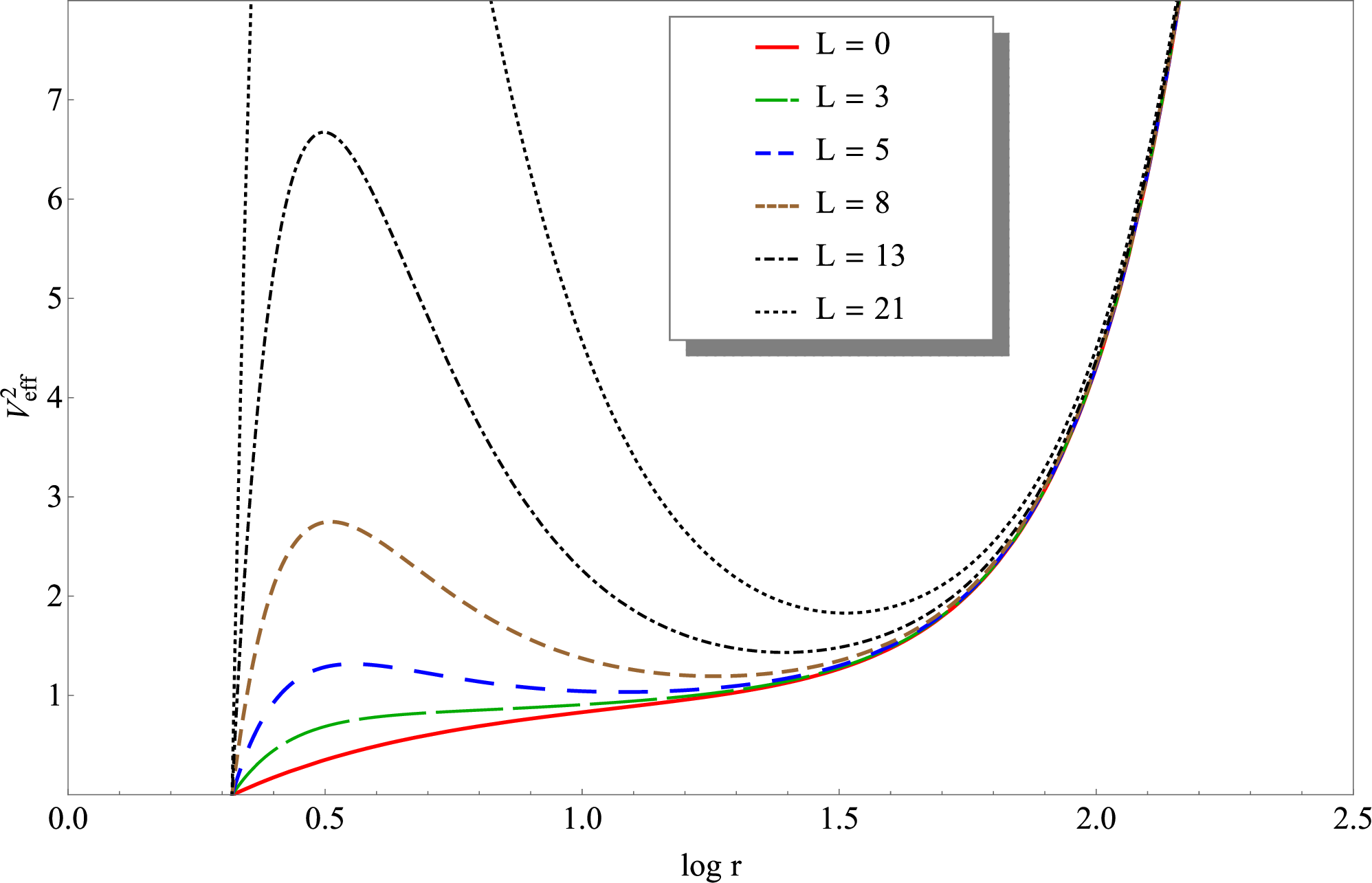}
	\includegraphics[width=0.48\textwidth]{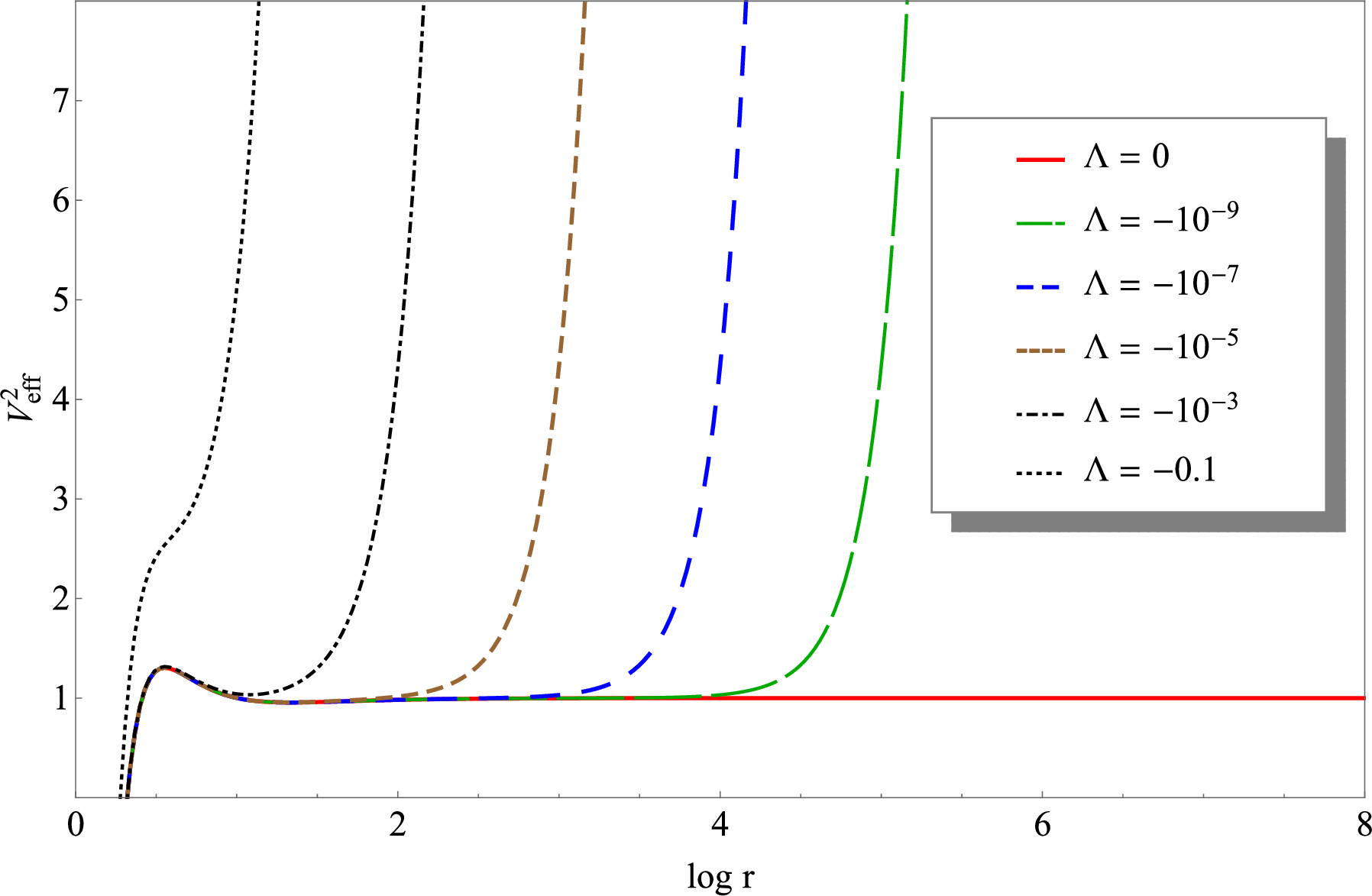}
	\caption{The behavior of the effective potential for a massive test particle in NC AdS Schwarzschild spacetime. \textbf{Left panel:} Different values of $L$, with fixed $\overline{\Theta} = 0.2$, $m = 1$, $\Lambda = -10^{-3}$ and $p_\varphi = 1$. \textbf{Right panel:} Different values of $\Lambda$, with fixed $m = 1$, $\overline{\Theta} = 0.2$, $p_\varphi = 1$ and $L = 5$.}
	\label{fig4}
\end{figure}

Figure \ref{fig4} shows the behavior of the non-commutative (NC) effective potential as a function of $r$, for varying $L$ (left panel) and varying $ \Lambda $ (right panel). As $L$ and $ \Lambda $ increase, both the maximum and minimum of the effective potential increase, making $ V_{\text{eff}}^{\text{NC}}$ less stable. For $ L < 3.58 $, these extrema disappear, indicating the absence of circular orbits. In the right panel, as $ \Lambda $ becomes more negative, two extrema are observed, and the radius of stable orbits (associated with the minimum) shifts toward the event horizon. When $ \Lambda = 0 $, the minimum disappears, and for $ \Lambda < -0.1 $, both extrema vanish, suggesting that stable circular orbits are no longer possible.

In the left panel of Figure \ref{fig4}, we present the behavior of the effective potential for a massive particle with constant values $L = 10$ and $\Lambda = -0.001$, for different values of $\overline{\Theta}$, and compare it to the commutative case. Note that for the NC BH, there are always two types of allowed orbits: the stable circular orbit, which corresponds to the minimum value of the effective potential at $ r_{\text{sta}}^{\text{NC}} > r_{\text{sta}}^{\text{C}} $, where $ (V_{\text{eff}}^{\text{NC}})_{\min} < (V_{\text{eff}}^{\text{C}})_{\min} $. This implies that the stable circular orbit in NC spacetime is more stable than in the commutative case. The other type is the unstable circular orbit, which corresponds to the maximum value of the effective potential, where in the NC case, the maximum is greater than in the commutative case, $ (V_{\text{eff}}^{\text{NC}})_{\max} > (V_{\text{eff}}^{\text{C}})_{\max} $. As observed, the curve in the NC case diverges near the origin, which is a manifestation of the existence of a minimal length scale that prevents probing distances smaller than $\sqrt{\Theta} $. It is also noted that in NC spacetime, there is always a minimum value for the orbital angular momentum $ (L > 3.58) $ and a maximum value for the cosmological constant $(\Lambda < -0.1)$, which are necessary conditions for the existence of stable and unstable circular orbits near the event horizon. The values $ L_{\text{crt}} = 3.58 $ and $ \Lambda_{\text{crt}} = -0.1 $ represent the critical points where the effective potential has an inflection point, known as the innermost stable circular orbit (ISCO). In this context, the NC geometry acts as a potential barrier near the event horizon \cite{abd}.

In general, a test particle is in a circular orbit when it moves in circular motion with a constant radius from the BH. Two types of circular orbits can be distinguished: stable circular orbits and unstable circular orbits, which can be determined from the condition:

\begin{equation}
	\frac{d}{dr}\left( V_{\text{eff}}^{\text{NC}}(r)\right)^2 = 0.
	\label{eq19}
\end{equation}

If $ V_{\text{eff}}^{\text{NC}}(r) $ has a minimum value, the circular orbit is stable, and if $ V_{\text{eff}}^{\text{NC}}(r) $ has a maximum value, the circular orbit is unstable. In the NC AdS Schwarzschild black hole, if the angular momentum satisfies $L > 3.58$ and $\Lambda < -0.1$, the NC effective potential develops extrema (maxima and minima), indicating the presence of unstable and stable circular orbits for a massive test particle. We report some numerical solutions to Equation \eqref{eq19} in Table \ref{tab1}, which shows the variation in the radius of stable and unstable circular orbits for different values of $ \overline{\Theta} $, with fixed parameters $ m = 1 $, $ L = 5 $, $ p_{\varphi} = 1 $, and $ \Lambda = -10^{-3} $.

\begin{table}[ht]
	\centering
	\caption{Numerical values of the unstable circular orbit $r_{\text{uns}}$ and stable circular orbit $r_{\text{sta}}$ for different values of the NC parameter $ \overline{\Theta} $, with fixed parameters $ m = 1 $, $ L = 5 $, $ p_{\varphi} = 1 $, and $ \Lambda = -10^{-3} $.}
	\label{tab1}
	\begin{tabular}{c c c c c c}
		\hline
		\textbf{$ \overline{\Theta} $} & 0 & 0.1 & 0.2 & 0.3 & 0.4 \\
		\hline
		\textbf{$ r_{\text{uns}} $} & 3.49578 & 3.54296 & 3.58522 & 3.62339 & 3.65814 \\
		\textbf{$ r_{\text{sta}} $} & 11.94138 & 11.9911 & 12.0401 & 12.0885 & 12.1364 \\ \hline
	\end{tabular}
\end{table}

\begin{figure}[ht]
	\centering
	\includegraphics[width=0.5\textwidth]{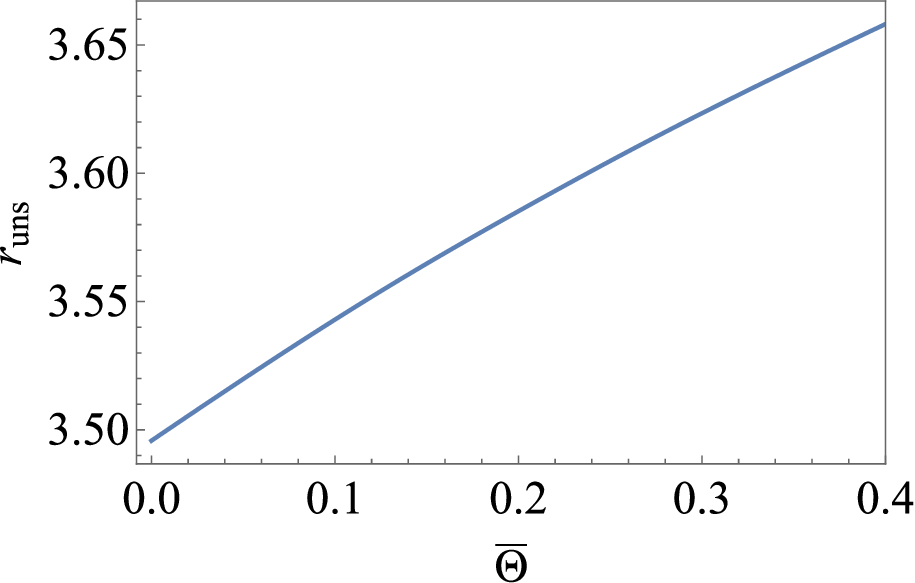}\hfill
	\includegraphics[width=0.5\textwidth]{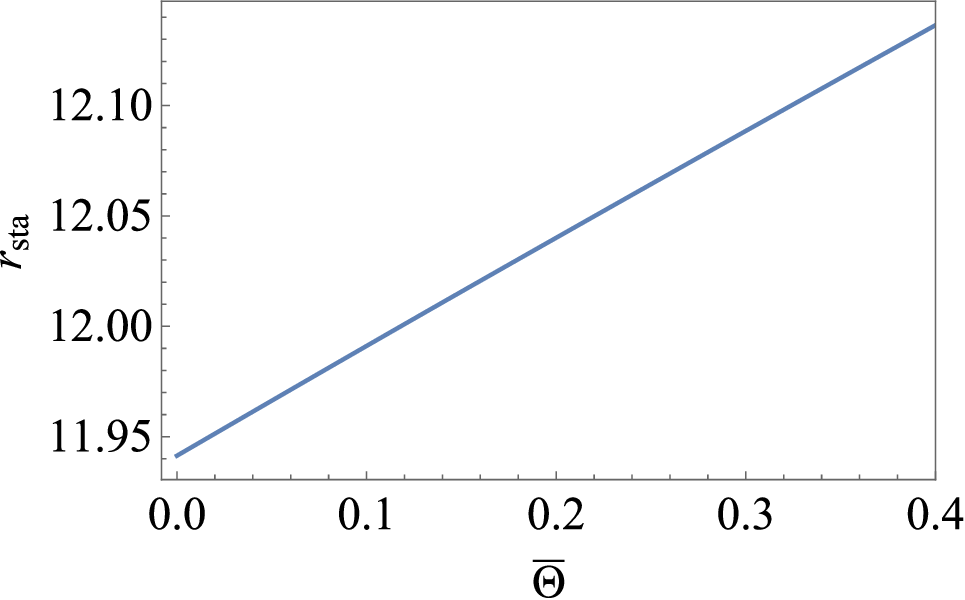}\hfill
	\caption{The behavior of the radii of circular orbits for a particle in NC spacetime. The panels show unstable and stable circular orbits as functions of $ \overline{\Theta} $ for fixed parameters: $ m = 1 $, $ L = 5 $, $ p_{\varphi} = 1 $, and $ \Lambda = -10^{-3} $.}
	\label{fig5}
\end{figure}

In Figure \ref{fig5}, we illustrate the behavior of the unstable and stable circular orbits of a massive test particle as functions of $ \overline{\Theta} $. In non-commutative spacetime, we observe that as the NC parameter increases, the radii of both unstable and stable circular orbits increase. This indicates that the massive test particle around the NC AdS Schwarzschild BH has more stable circular orbits.


\section{Orbital Equation and Advance of the Perihelion}\label{sec:NCOEAP}

We now derive the geodesic motion of test particles around a NC AdS Schwarzschild BH for the metric given by Eq. \ref{eq9}. The equation of motion is obtained using Eq. \ref{eq16b} and Eq. \ref{eq17}, with the change of variable $ u = \frac{1}{r} $, giving:
\begin{align}
	\left(\frac{du}{d\varphi}\right)^2 &= \frac{(E^2 - 1)}{L^2} + \frac{2m}{L^2} u - u^2 + 2m u^3 + \frac{\Lambda}{3} + \frac{\Lambda}{3 L^2 u^2} \notag \\
	&+ \overline{\Theta} p_\varphi \left\{ \left(\frac{2}{L^2} - \frac{2E^2}{L^2} - \frac{2}{3}\Lambda\right) u - \frac{3m}{L^2} u^2 + u^3 - m u^4 - \frac{\Lambda}{L^2 u} \right\} + \mathcal{O}(\overline{\Theta}^2).
	\label{eq20}
\end{align}

We differentiate this equation with respect to $\varphi$ to obtain the NC geodesic equation for a massive test particle to first order in the NC parameter $ \overline{\Theta} $:
\begin{equation}
	\frac{d^2 u}{d\varphi^2} + u = \frac{m}{L^2} + 3m u^2 - \frac{\Lambda}{3L^2 u^3} + \overline{\Theta} p_\varphi \left\{ \left(\frac{1}{L^2} - \frac{E^2}{L^2} - \frac{\Lambda}{3}\right) - \frac{3m}{L^2} u + \frac{3}{2} u^2 - 2m u^3 + \frac{\Lambda}{2L^2 u^2} \right\}.
	\label{eq21}
\end{equation}

Note that the last term on the right-hand side represents the corrections induced by NC effects up to the first order in $ \overline{\Theta} $. In the limit $ \overline{\Theta} \to 0 $, this term tends to zero, recovering the orbit analyzed in \cite{Ads.Sch}. Equation \eqref{eq21} is difficult to solve analytically, so we solve it numerically to investigate the structure of time-like geodesics and how the NC parameter of spacetime influences the time-like geodesics in the NC AdS Schwarzschild spacetime.

We plot the geodesic equation (Eq. \eqref{eq21}) for a massive test particle around a NC AdS-Schwarzschild spacetime, considering different values of $ L $ and $ E $, with the BH mass fixed at $m = \frac{3}{14}$. We compare the commutative case ($ \overline{\Theta} = 0 $) with the NC one ($ \overline{\Theta} = 0.1 $), and also compare Schwarzschild spacetime ($ \Lambda = 0 $) with AdS Schwarzschild spacetime ($ \Lambda = -0.001 $).

\begin{figure}[ht]
	\centering
	\includegraphics[width=0.548\textwidth]{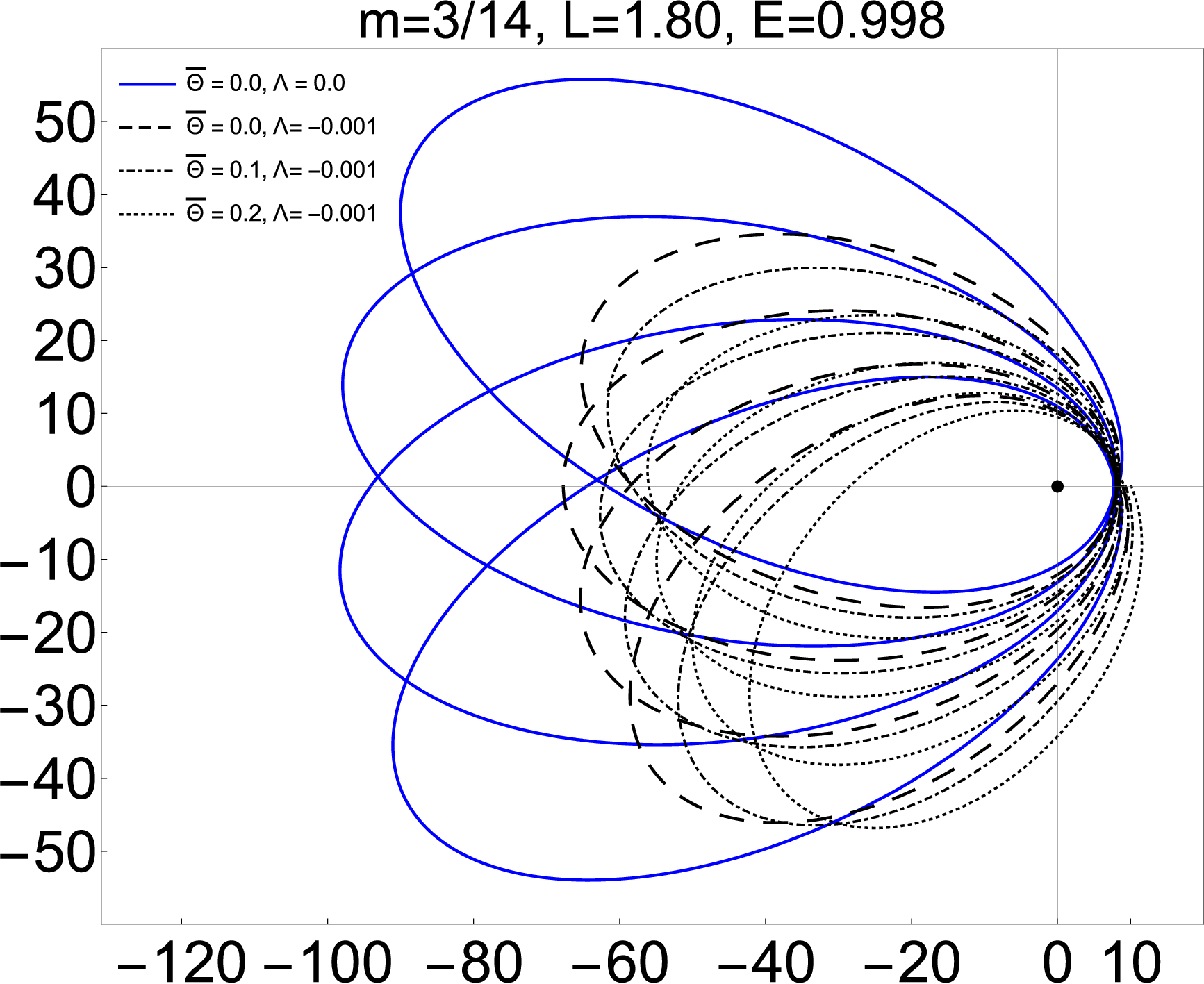}\hfill
	\includegraphics[width=0.449\textwidth]{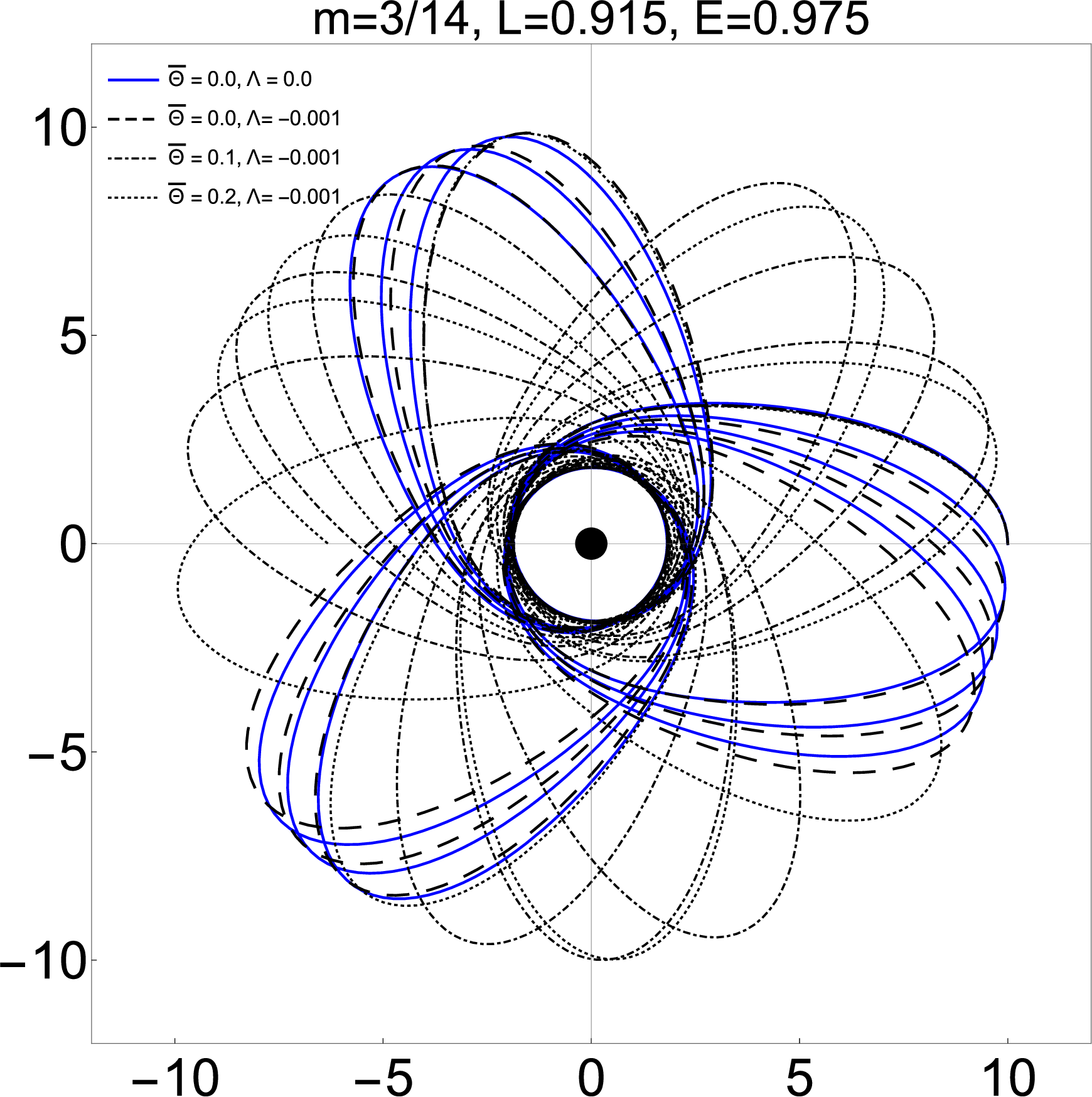}
	\caption{Time-like geodesics for a massive test particle around a NC AdS Schwarzschild BH, for different values of $ \overline{\Theta} $ and $ \Lambda $.}
	\label{fig6}
\end{figure}

In Fig. \ref{fig6}, we describe the motion of a massive test particle around the NC AdS Schwarzschild BH, depending on the choice of relevant parameters such as mass, angular momentum, cosmological constant, energy, and the NC parameter. This analysis takes into account the effective potential at the limiting maximum and minimum values. We observe that the orbital plane remains invariant and maintains its shape, but it precesses due to the presence of the NC parameter. This precession is significant when compared to the precession rate in the commutative case in the presence of the cosmological constant. Moreover, for fixed values of the cosmological constant, the effect of the NC parameter increases the semi-major axis of the particle's orbit, making the orbit more stable, as seen in both the left and right panels.


\subsection{Advance of the Perihelion}\label{sec:advance_perihelion}

To obtain the expression for the periastron advance of the orbit, we rewrite the orbital equation \eqref{eq21} in a perturbed form based on the Keplerian trajectory equation \cite{kepler}:
\begin{equation}
	\frac{d^2 u}{d\varphi^2} + u = \frac{m}{L^2} + \frac{g(u)}{L^2},
	\label{eq22}
\end{equation}
where 
\begin{equation}
	\frac{g(u)}{L^2} = 3m u^2 - \frac{\Lambda}{3 L^2 u^3} + \overline{\Theta} p_\varphi \left\{ \left(\frac{1}{L^2} - \frac{E^2}{L^2} - \frac{\Lambda}{3}\right) - \frac{3m}{L^2} u + \frac{3}{2} u^2 - 2m u^3 + \frac{\Lambda}{2L^2 u^2} \right\}.
\end{equation}

Following the steps outlined in \cite{orbital1, abd}, we obtain the angular deviation after one complete revolution, defined as:
\begin{equation}
	\Delta\varphi = \frac{\pi}{L^2}\Bigg| \frac{d g(u)}{d u} \Bigg|_{u = \frac{1}{b}},
	\label{eq24}
\end{equation}
where $ b = a(1 - e^2) $, with $e$ and $a$ representing the orbit's eccentricity and semi-major axis, respectively. We also have $ m = \frac{GM}{c^2} $, $ p_{\varphi} = M V_{\varphi} $, and $ L^2 = \frac{GM}{c^2} a (1 - e^2) $. Hence, the precession per revolution is given by:
\begin{equation}
	\Delta\varphi = \frac{6\pi GM}{c^2 \alpha (1 - e^2)} + \frac{\pi c^2 \alpha^3 (1 - e^2)^3}{GM} \Lambda - \pi \overline{\Theta} M V_\varphi \left\{ \frac{6GM}{c^2 \alpha^2 (1 - e^2)^2} + \frac{c^2 \alpha^2 (1 - e^2)^2}{GM} \Lambda \right\}.
	\label{eq25}
\end{equation}

The first term represents the prediction of general relativity, the second term corresponds to the contribution of the negative cosmological constant \cite{rindler}, and the last term represents the first-order correction from the NC parameter, which accounts for the new advance due to NC effects. As we can see, when $ \overline{\Theta} = 0 $, we recover the commutative advance \cite{rindler}. 

For a numerical application, we use the example of the planet Mercury \cite{V_phi, Lambda-value, Values}:
\begin{equation}
	\Delta\varphi = 2\pi \times (7.98744 \times 10^{-8}) + 2\pi \times (5.77633 \times 10^{-24}) - 2\pi \times (6.93249 \times 10^{18}) \times \overline{\Theta}.
\end{equation}

The first term represents the prediction of general relativity, the second term corresponds to the contribution of the cosmological constant, and the third term accounts for the NC corrections. The observed perihelion shift of Mercury is given by \cite{obs}:
\begin{equation}
	\Delta\varphi_{\text{obs}} = 2\pi \times (7.98734 \pm 0.00037) \times 10^{-8} \, \text{rad/rev}.
\end{equation}

We can now define a lower bound for the NC parameter $ \overline{\Theta} $ using the following condition:
\begin{align}
	\left| \delta\varphi_{\text{NC}} \right| &\leq \left| \Delta\varphi_{\text{th}} - \Delta\varphi_{\text{obs}} \right| \approx 2\pi \times (2.7) \times 10^{-12}.
\end{align}

From this, we find:
\begin{equation}
	\overline{\Theta} \leq 3.89369 \times 10^{-30} \, \text{s} \cdot \text{kg}^{-1}.
\end{equation}
or equivalently,
\begin{equation}
	\sqrt{\hbar \overline{\Theta}} = \sqrt{\Theta} \leq 2.02637 \times 10^{-32} \, \text{m} \sim 10^3 l_P,
	\label{eq30}
\end{equation}
where $ l_P $ is the Planck length. 

Note that this result is approximately the same as the one obtained in reference \cite{kepler}, where the authors used Newtonian mechanics in a NC flat spacetime. We also observe a difference of order $ 10^{-1} $ with the result from \cite{abd, abdellah3}, which was obtained using NC gauge theory in Schwarzschild spacetime. This difference arises because we included the cosmological constant $ \Lambda $. For a given value of $ \Lambda $, we can obtain the same result as we did here. 

Thus, the planetary system is very sensitive to the NC parameter, implying that the NC parameter plays the role of a fundamental constant in describing the fine structure of spacetime. A small change in $ \Theta $ results in a significant change in the system on a large scale. From equation \eqref{eq30}, we notice that the minimum NC parameter also has a minimum value corresponding to the Planck scale $ l_P $.

Finally, in natural units, we can obtain the upper bound for the energy:
\begin{equation}
	\frac{1}{\sqrt{\Theta}} \geq 9.76633 \times 10^{15} \, \text{GeV} \approx 10^3 E_P,
\end{equation}
indicating that the properties of the NC geometry manifest at high-energy scales.


\subsection{Second-order Correction to the Advance of the Perihelion}
\label{sec:second_order_correction}

Using the definition in \eqref{eq7}, we can determine the components of the deformed metric $ \hat{g}_{\mu\nu}(x, \Theta) $ in Eqs. \eqref{eq10}, \eqref{eq11}, \eqref{eq12}, and \eqref{eq13}, up to the second order in $ \overline{\Theta} $. These components are given by:

\begin{align}
	-\hat{g}_{tt} &= \left(1 - \frac{2m}{r} - \frac{\Lambda}{3} r^2 \right) - \overline{\Theta} p_\varphi \left\{\frac{m}{r^2} - \frac{\Lambda}{3} r \right\} - \overline{\Theta}^2 \frac{p_\varphi^2}{2} \left\{\frac{m}{r^3} + \frac{\Lambda}{6} \right\} + \mathcal{O}(\overline{\Theta}^3), 
	\label{eq32}\\
	\hat{g}_{rr} &= \left(1 - \frac{2m}{r} - \frac{\Lambda}{3} r^2 \right)^{-1} + \overline{\Theta} p_\varphi \left\{\left(\frac{m}{r^2} - \frac{\Lambda}{3} r\right) \left(1 - \frac{2m}{r} - \frac{\Lambda}{3} r^2 \right)^{-2} \right\} \notag \\
	&\quad + \overline{\Theta}^2 \frac{p_\varphi^2}{4} \left\{\frac{9 \left(-6m + 12mr^2\Lambda - r^3\Lambda - r^5\Lambda^2 \right)}{\left(6m - 3r + r^3\Lambda\right)^3} \right\} + \mathcal{O}(\overline{\Theta}^3),
	\label{eq33}\\
	\hat{g}_{\theta\theta} &= r^2 - r \overline{\Theta} p_\varphi + \overline{\Theta}^2 \frac{p_\varphi^2}{4} + \mathcal{O}(\overline{\Theta}^3),
	\label{eq34}\\
	\hat{g}_{\varphi\varphi} &= r^2 \sin^2 \theta - r \overline{\Theta} p_\varphi \sin^2 \theta + \overline{\Theta}^2 \frac{p_\varphi^2}{4} \sin^2 \theta + \mathcal{O}(\overline{\Theta}^3).
	\label{eq35}
\end{align}

Following the same steps as before, we can derive the explicit expression for the angular deviation in NC spacetime, with a correction up to the second order in $\overline{\Theta}$:
\begin{align}
	\Delta\varphi &= \frac{6\pi GM}{c^2 \alpha (1 - e^2)} + \frac{\pi c^2 \alpha^3 (1 - e^2)^3}{GM} \Lambda - \pi \overline{\Theta} M V_\varphi \left\{ \frac{6GM}{c^2 \alpha^2 (1 - e^2)^2} + \frac{c^2 \alpha^2 (1 - e^2)^2}{GM} \Lambda \right\} \notag \\
	&\quad + \pi \overline{\Theta}^2 M^2 V_\varphi^2 \left\{ \frac{3}{\alpha^2 (1 - e^2)^2} - \frac{3c^2}{2GM \alpha (1 - e^2)} + \frac{3c^2 E^2}{2GM \alpha (1 - e^2)} + \frac{\Lambda}{2} \right\}.
	\label{eq36}
\end{align}

For the case of Mercury \cite{V_phi, Lambda-value}, we obtain from \eqref{eq36} a new limit bound for the NC parameter \( \Theta \):
\begin{equation}
	\sqrt{\Theta} \leq 4.07787 \times 10^{-33} \, \text{m} \sim 10^2 l_P,
\end{equation}
where $ l_P $ is the Planck length.

As shown, the second-order correction provides a more precise estimate for the NC parameter $ \sqrt{\Theta} $, which is on the order of $ 10^2 l_P $. This suggests that higher-order corrections should be computed for even more accurate estimations. However, only a few studies have constrained the NC parameter through experimental tests of general relativity. For example, the four classical tests \cite{abdellah3} and gravitational wave studies \cite{ThetaGW} suggest that the NC parameter $ \sqrt{\Theta} $ is expected to be bounded between $ 10^{-34} \, \text{m} $ and $ 10^{-35} \, \text{m} $. Additionally, studies on the lower bound of $\Theta$ through BH thermodynamics provide promising estimates at the Planck scale, which are on the order of $ 10^{-35} \, \text{m} $. For instance, Refs. \cite{abdellah6, abdellah4, abdellah5} use the NC gauge theory of gravity to obtain such estimates. Some papers have also employed NC coherent states of matter in the study of BH thermodynamics \cite{piero1, nicolini2, alavi}, where their analysis suggests that the NC parameter $ \sqrt{\Theta} $ is on the order of $10^{-1} l_P$. Furthermore, our results reaffirm that the NC nature of spacetime is manifested at scales below the Planck length.



\section{Conclusion}\label{sec:conl}

Using the star product method (Bopp's shift method), we have determined the NC corrections to the AdS Schwarzschild solution up to the first and second orders in the NC parameter $\Theta$. We have shown that the event horizon in the NC AdS Schwarzschild BH is larger than in the commutative case, meaning that the gravitational force at the event horizon of a NC BH is stronger compared to that of a commutative BH.

The effective potential for particles in the NC AdS Schwarzschild BH has been calculated. Through detailed analysis, we demonstrated that massive test particles orbiting a NC AdS Schwarzschild BH have more stable circular orbits. Additionally, we have considered the periastron advance in NC geometry and calculated the NC perihelion $ \Delta \varphi $ for the planet Mercury. Our results, based on experimental data, indicate that the lower bound of the NC parameter $ \sqrt{\Theta} $ is on the order of $ 10^{-32} \, \text{m} $, which is larger than the Planck length. By extending the correction to second order in NC parameter, we obtained a new lower bound for $ \sqrt{\Theta} $ on the order of $ 10^{-33} \, \text{m} $. These estimates suggest that the NC nature of spacetime becomes significant before reaching the Planck scale.



\section*{Acknowledgments}
This work is supported by PRFU research project B00L02UN050120230003, Univ. Batna 1, Algeria.

\bibliographystyle{unsrt}  
\bibliography{references}

\end{document}